\documentclass{emulateapj}
\usepackage{amsmath}
\usepackage{mathptmx}
\newcommand{\unitspace}{\ensuremath{\,}}
\newcommand{\usp}{\unitspace}
\newcommand{\numberspace}{\ensuremath{\;}}
\newcommand{\nsp}{\numberspace}
\newcommand{\unitstyle}[1]{\ensuremath{\mathrm{#1}}}
\newcommand{\power}[2]{\ensuremath{{#1}^{#2}}}

\newcommand{\centi}{\unitstyle{c}}
\newcommand{\kilo}{\unitstyle{k}}

\newcommand{\meter}{\unitstyle{m}}
\newcommand{\gram}{\unitstyle{g}}
\newcommand{\second}{\unitstyle{s}}

\newcommand{\cm}{\centi\meter}
\newcommand{\Kelvin}{\unitstyle{K}}
\newcommand{\K}{\Kelvin} 

\newcommand{\grampercc}{\gram\usp\power{\cm}{-3}} 
\newcommand{\cmpersec}{\cm\usp\power{\second}{-1}} 

\newcommand{\Msun}{\ensuremath{M_\odot}}

\newcommand{\km}{\kilo\meter}   

\newcommand{\ee}[1]{\ensuremath{\times 10^{#1}}}

\newcommand{\code}[1]{\textsc{#1}}

\newcommand{\nuclei}[2]{\ensuremath{\mathrm{^{#1}#2}}}

\newcommand{\carbon}[1][12]{\nuclei{#1}{C}}

\newcommand{\oxygen}[1][16]{\nuclei{#1}{O}}

\newcommand{\neon}[1][20]{\nuclei{#1}{Ne}}
\newcommand{\sodium}[1][23]{\nuclei{#1}{Na}}

\newcommand{\silicon}[1][28]{\nuclei{#1}{Si}}

\newcommand{\nickel}[1][58]{\nuclei{#1}{Ni}}

\newcommand{\krypton}[1][84]{\nuclei{#1}{Kr}}

\newcommand{\SNeIa}{SNe~Ia}
\newcommand{\SNIa}{SN~Ia}

\begin{document}
\shorttitle{Nucleosynthetic Signature of Edge-lit Detonation }
\shortauthors{Chamulak et al.}  
\title{Asymmetry and the Nucleosynthetic Signature of Nearly Edge-Lit Detonation in White Dwarf Cores} 
\author{David A. Chamulak\altaffilmark{1,7}, Casey
  A. Meakin\altaffilmark{2,3}, Ivo R. Seitenzahl\altaffilmark{4}, James
  W. Truran\altaffilmark{1,5,6,7}} \altaffiltext{1}{Argonne National
  Laboratory, Argonne, IL} \altaffiltext{2}{Steward Observatory,
  University of Arizona, Tucson, AZ} \altaffiltext{3}{Los Alamos National
  Laboratory, Los Alamos, NM} 
  \altaffiltext{4}{Max Planck Institute for Astrophysics, Garching, Germany}
  \altaffiltext{5}{Department of Astronomy and Astrophysics, University
  of Chicago, Chicago, IL} \altaffiltext{6}{Center for Astrophysical
  Thermonuclear Flashes, University of Chicago, Chicago, IL}
\altaffiltext{7}{Joint Institute for Nuclear Astrophysics}
\email{dchamulak@anl.gov}

\begin{abstract}
  Most of the leading explosion scenarios for Type Ia supernovae
  involve the nuclear incineration of a white dwarf star through a
  detonation wave.  Several scenarios have been proposed as to how
  this detonation may actually occur, but the exact mechanism and
  environment in which it takes place remain unknown.  We explore the
  effects of an off-center initiated detonation on the spatial
  distribution of the nucleosynthetic yield products in a toy model --
  a pre-expanded near Chandrasekhar-mass white dwarf.  We find that a
  single-point near edge-lit detonation results in asymmetries in the
  density and thermal profiles, notably the expansion timescale,
  throughout the supernova ejecta.  We demonstrate that this asymmetry of the
  thermodynamic trajectories should be common to off-center
  detonations where a small amount of the star is burned prior to
  detonation. The sensitivity of the yields
  on the expansion timescale results in an asymmetric distribution of
  the elements synthesized as reaction products.  We tabulate the
  shift in the center of mass of the various elements produced in our model supernova and
  find an odd-even pattern for elements past silicon. Our calculations show that off-center single-point
  detonations in carbon-oxygen white dwarfs are marked by
  significant composition asymmetries in their remnants
  which bear potentially observable
  signatures in both velocity and coordinate space, including an
  elemental nickel mass fraction which varies by a factor of two to
  three from one side of the remnant to the other.
\end{abstract}


\section{Introduction}\label{sec:introduction}
\label{s:introduction}
The fact that their peak absolute magnitude is correlated with the
width of the light curve has allowed Type Ia supernovae (\SNeIa ) to
be used as standard candles in determining cosmological
distances. Despite this widespread use of \SNeIa\ as standard
candles, many problems in describing how the explosion of the star
happens still remain.

A number of models have been proposed to explain \SNeIa\ \citep[for a review, see][]{hillebrandt.niemeyer:type,podsiadlowski2008}.
Two promising candidates for an explosion mechanism are the
sub-Chandrasekhar double detonation model \citep{woosley1994,fink2010}
and the gravitational confined detonation (GCD) model
\citep{plewa.calder.ea:type}. In both models, most if not all of the
nuclear burning occurs in a detonation wave. In the sub-Chandrasekhar
model a layer of helium is deposited on the surface of a white
dwarf. The helium layer detonates resulting in a shock wave traveling
around the surface of the white dwarf.  When the shock wave converges
at the antipode, a detonation is thought to be triggered off center in the carbon-oxygen core
\citep{woosley1994,fink2010,sim2010}. The nucleosynthetic
yield is set predominantly by the mass of the carbon-oxygen white dwarf
core and the mass of the helium layer \citep{woosley1994,fink2010}.

In the GCD model, the carbon burning runaway within the convective
core of a near Chandrasekhar-mass white dwarf is postulated to occur in a small
region displaced from the stellar center that becomes a highly
buoyant flame bubble and quickly rises to the stellar surface after
burning only a few percent of the star during its ascent
\citep[e.g.][]{plewa.calder.ea:type, Livne2005On-the-Sensitiv}. When the buoyant ash, as
well as unburned material pushed ahead of the rising flame bubble,
erupts forth from the stellar core it is largely confined to the
surface of the white dwarf by gravity. It then becomes a strong surface
flow that sweeps completely over the star, eventually converging in a
region opposite to the breakout location. Although the details of the detonation initiation process is still a
topic of active research \citep[see e.g.,][]{Ropke:2007fu,seitenzahl2009b,seitenzahl2009c} the
high temperatures and densities reached within the converging surface
flow are strong indicators that a detonation is likely to be
triggered.

The resulting GCD nucleosynthetic yield consists almost entirely of
detonation burning products and depends on how much the star has
expanded by the time the detonation initiates.  More highly expanded
(hence lower density) cores at detonation result in a smaller fraction
of Fe peak nuclei, less \nickel[56], and consequently a larger
fraction of intermediate mass elements (IMEs) due to incomplete
relaxation to nuclear statistical equilibrium (NSE).  Therefore, lower
luminosity (less \nickel[56] producing) explosions are accompanied by
a larger yield of IMEs, as has been observed \citep[see e.g.,][]{mazzali2007}.

The expansion of the star prior to detonation, in a GCD scenario, results from the work
done by the rising flame bubble.  Deflagrations that burn
more mass prior to reaching the stellar surface excite higher
amplitude pulsations and hence more-expanded stars at the time of
detonation. It has been found that the expansion of the star due to
the deflagration is very well represented by the fundamental radial
pulsation mode of the underlying white dwarf \citep[see Fig. 2 and
Fig. 12 in][]{meakin2009}.  Therefore, while it is essential to
understand the nature of the deflagration so as to better understand
the mapping between initial conditions and final outcomes, the range
of outcomes due to the deflagration can be parameterized by the
pulsation amplitude, resulting in a one parameter family of
models. While the phase of the pulsation at the time of detonation is
an additional parameter, it plays a lesser role since the pulsation
period is longer then the detonation crossing time.

In this Paper we explore the nucleosynthetic yields that result from
an edge-lit detonation of a pre-expanded near Chandrasekhar-mass white
dwarf core out of hydrostatic equilibrium. For our purposes we define edge-lit 
detonation to be an off-center detonation where very little of the star has burned 
before hand.
This is a toy-model which neglects the effect the deflagration has on the nucleosynthesis.
However, due to the small amount of mass burned in the deflagration it
captures most of the essential features of the GCD model. The physical
mechanism that gives rise to the abundance asymmetries in the ejecta
will furthermore manifest itself in other off-center detonation
models, such as the sub-Chandrasekhar double detonation models.

In Section~\ref{s:numerical-model} we describe the explosion model and
our computational method. Section~\ref{s:results} is comprised of
three parts. First, we discuss how the dynamics of an asymmetric
detonation affect the hydrodynamic profile of the \SNIa\ remnant. We
then consider how variations in the thermodynamic trajectories of
detonated material affect the resulting nucleosynthesis.  Finally, we
present the elemental yields found for our explosion model and
quantify the asymmetric distribution thereof in terms of center of
mass offsets, which we provide in tabulated form. We also show that
there is an asymmetric distribution in velocity space for elemental nickel. We then
conclude in Section~\ref{s:discussion}, with a discussion of how our
results relate to previous work and how our work can be improved.

\section{The Explosion Model}\label{s:numerical-model}

The explosion model discussed in this paper involves the 
detonation  of a cold (T=$4\ee{7}\nsp\K$) white
dwarf of mass $1.365\Msun$, comprised of 50\% \carbon\ and 50\% \oxygen\
by mass, which has been expanded according to its fundamental
radial-pulsation mode by such an amplitude that it has a central
density of $10^8\nsp\grampercc$.   A detonation is initiated in this expanded white dwarf at a radial location
of $r=2\times 10^7$cm, where the density of the white dwarf is 
$\rho= 10^7\grampercc$, by heating a small spherical volume with 
radius $r_{\rm ini}\sim 4$ km to high temperature $T_{\rm ini}\sim10^9$K.  This 
heating immediately triggered a detonation which propagates away from the 
point of initiation.

The subsequent reactive-hydrodynamic evolution of the detonated white dwarf was conducted using 
the FLASH code \citep{fryxell.ea:flash}. The code framework and the included physics
is identical to that described in \citet{meakin2009}.  The star was set up on a
Cartesian grid in 2D with cylindrical symmetry.  An effective adaptive mesh refinement 
resolution of 1\nsp\km\ is used. The detonation was initiated along the symmetry
axis, which is the only natural way to capture its evolution in this geometry. 
Due to the cylindrical symmetry of the physical scenario and the lack of large scale instability in the detonation front
this setup gives the same results as a full 3D calculation of the same problem 
(see \citet[][]{meakin2009} which provides an in depth description of the 
detonation phase, and a comparison between 2D and 3D).  

Energy release and bulk compositional evolution due to nuclear burning is coupled to the hydrodynamic flow by advancing a 
system of progress variables which represent three stages of burning: (1) the burning 
of carbon and oxygen to silicon group elements, (2) the relaxation of the silicon group elements
to NSE, and (3) the evolution of material which has already relaxed entirely to NSE, including its
adjustment to changing density and temperature and neutronization due to weak interactions 
among the Fe peak elements.  Additional details can be found in \citet{Calder2007Capturing-the-F,meakin2009}. 

The pre-expansion of the white dwarf prior to detonation initiation results in $\sim
1\Msun$ of high density material ($\rho>10^7\nsp\grampercc$), which
burns to NSE in the detonation (primarily as \nickel[56]). The
remaining $\sim0.37\Msun$ of material burns to IMEs, e.g. Si, S, Ca,
and Ar, resulting in only a small amount ($<0.04\Msun$) of unburned
C/O in the outermost layers of the remnant. 

Detailed yields are calculated by post processing Lagrangian tracer particles included
in the explosion calculation and are the primary focus of this
paper. We recorded the time history of 10$^4$ particles which were
initialized to evenly sample the initial mass of the white dwarf. Our reaction network incorporates 
493 nuclides from n to \krypton[86] (Table~\ref{t:network}). We use the reaction rates from the Joint
Institute for Nuclear Astrophysics \code{reaclib}
Database\footnote{\url{http://groups.nscl.msu.edu/jina/reaclib/db/}}
\citep[][and references therein]{cyburt2010}; the light-element rates
are mostly experimental and are from compilations such as
\citet{caughlan88:_therm} and \citet{Iliadis2001Proton-induced-}. Weak
reaction rates are taken from \citet{Fuller1982Stellar-weak-in} and
\citet{langanke.martinez-pinedo:weak}. Screening is incorporated using
the formalism of \citet{graboske.dewitt.ea:screening}.

\begin{deluxetable}{crcrcrcr}
  \tablecaption{493-Nuclide Reaction Network} \tablehead{\colhead{El.}
    & \colhead{$A$} & \colhead{El.} & \colhead{$A$} & \colhead{El.} &
    \colhead{$A$} & \colhead{El.} & \colhead{$A$}} \startdata
  n & & & & & & & \\
  H	&	1--3	&	Ne	&	17--28	&	K	&	35--46	&	Ni	&	50--73	\\
  He	&	3--4	&	Na	&	20--31	&	Ca	&	35--53	&	Cu	&	54--70	\\
  Li	&	6--8	&	Mg	&	20--33	&	Sc	&	40--53	&	Zn	&	55--72	\\
  Be	&	7, 9--11	&	Al	&	22--35	&	Ti	&	39--55	&	Ga	&	58--73	\\
  B	&	8, 10-14	&	Si	&	22--38	&	V 	&	43--57	&	Ge	&	59--76	\\
  C	&	9--16	&	P	&	26--40 	&	Cr 	&	43--60	&	As	&	62--76	\\
  N	&	12--20	&	S	&	27--42	&	Mn	&	46--63	&	Se	&	62--82	\\
  O	&	13--20	&	Cl	&	31--44	&	Fe	&	46--66	&	Br	&	71--81	\\
  F	&	15--24	&	Ar	&	31--47	&	Co	&	50--67	&	Kr	&	71--86	\\
  \enddata
  \label{t:network}
\end{deluxetable}

\section{Results}
\label{s:results}

\subsection{Explosion Dynamics and Remnant Asymmetry}

\label{s:explosionDynamics}

The detonation propagates from the point of initiation at nearly
the Chapman--Jouguet (CJ) speed, $D_{CJ}\sim 1.2\ee{9}\nsp\cmpersec$
at a Mach number of $M_{CJ}\sim D_{CJ}/c_s \sim 3.4$.  Because of the
weak upstream density dependence of the detonation speed under these
conditions, the detonation front remains very nearly spherical in
shape as it engulfs the star.  The total time required for its passage
across the expanded white dwarf core is $t_{cross}\sim 2 r/D_{CJ}\sim
0.5\nsp\second$.  This is followed by a period of $\sim
0.5\nsp\second$ in which the pressure forces drive the completely
incinerated remnant into a homologous expansion, characterized by a
purely radial expansion velocity profile with an expansion rate
proportional to the radial position $v \propto r$.  After only a few
seconds, the remnant is expanding ballistically, and the total energy
budget is dominated by the kinetic energy. The homologous velocity
profile results in a self-similar density profile which persists until
the remnant begins to interact with the interstellar medium.

\begin{figure}[htbp]
  \includegraphics[width= 3.5in]{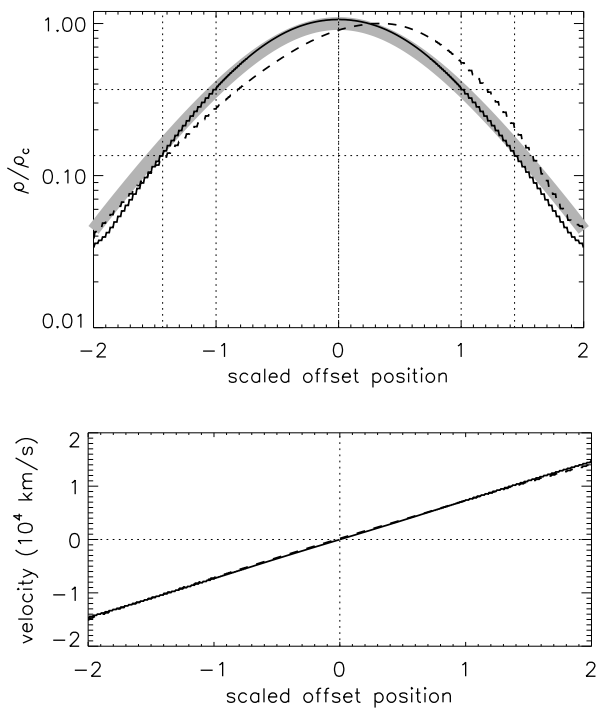}
  \caption{Late time (t $>$ 3 s) density and velocity profiles for the
    post-detonation state.  The density is scaled by the peak value
    and the position is scaled by the density e-folding distance in
    the equatorial direction. The thick gray line shows the scaled
    density profile of the initial white dwarf model, while the post-detonation
    state model is shown by thick black lines for lines through the
    equator (solid) and through the poles (dashed).}
  \label{f:detprofile}
\end{figure}

The expanding remnant resulting from the detonation is marked by
significant asymmetry. The late time density profiles along the
symmetry axis and the equator are shown in Fig~\ref{f:detprofile},
scaled by the peak density in the remnant $\rho_c$. The initial,
spherically symmetric white dwarf density profile is also shown for
comparison. It can be seen that the density peak is shifted into the
hemisphere in which the detonation was initiated, $y>0$ in this case,
resulting in a steeper density gradient in this hemisphere.  This is
in agreement with the series of GCD simulations described in
\citet{meakin2009}(see their Fig. 9 - 11) which show that the density
isocontours in the remnant are well described by concentric circles
that have centers offset from the initial stellar center.  The density
isocontours were found to have larger offsets at higher densities,
with the largest offset centered on the peak density in the remnant.
{\em In all cases the density peaks on the side of the
  remnant where the detonation originated and has a steeper density gradient in that region.}

Unlike the density profile, the velocity profile (also shown in
Fig.~\ref{f:detprofile}) does not show an asymmetry, but is everywhere
radially directed and spherically symmetric. This leads to an
asymmetry in the density as a function of expansion velocity, which is
likely to result in a viewing angle dependence for the light curve and
the spectral signature. Related composition asymmetries, discussed in
\S\S\ref{s:NucDepend} and \ref{s:Phenom} below, also contribute
to observable asymmetries and viewing angle dependencies.

\begin{figure*}[htbp]
  \includegraphics[width= 3.5in]{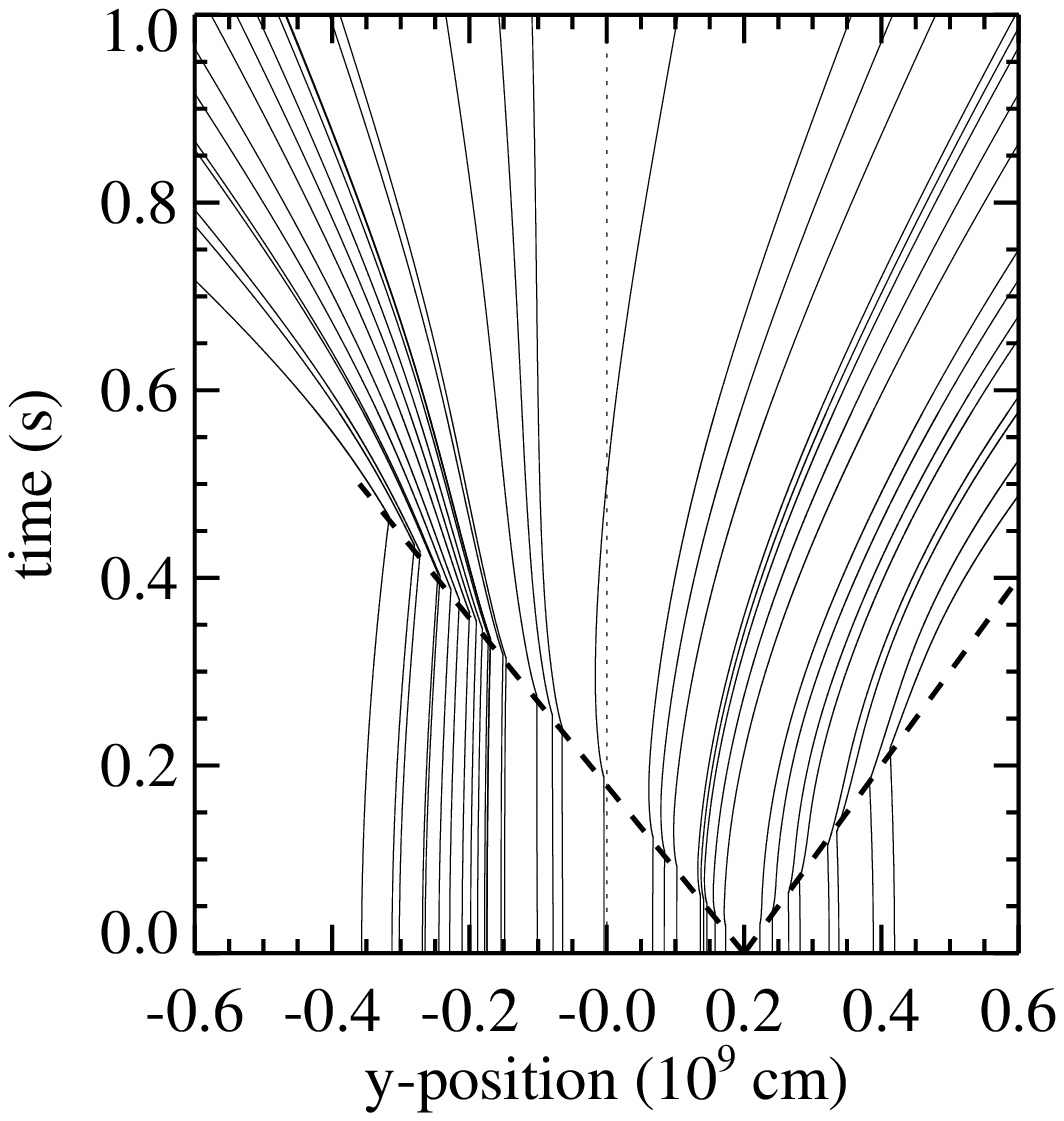}
  \includegraphics[width= 3.5in]{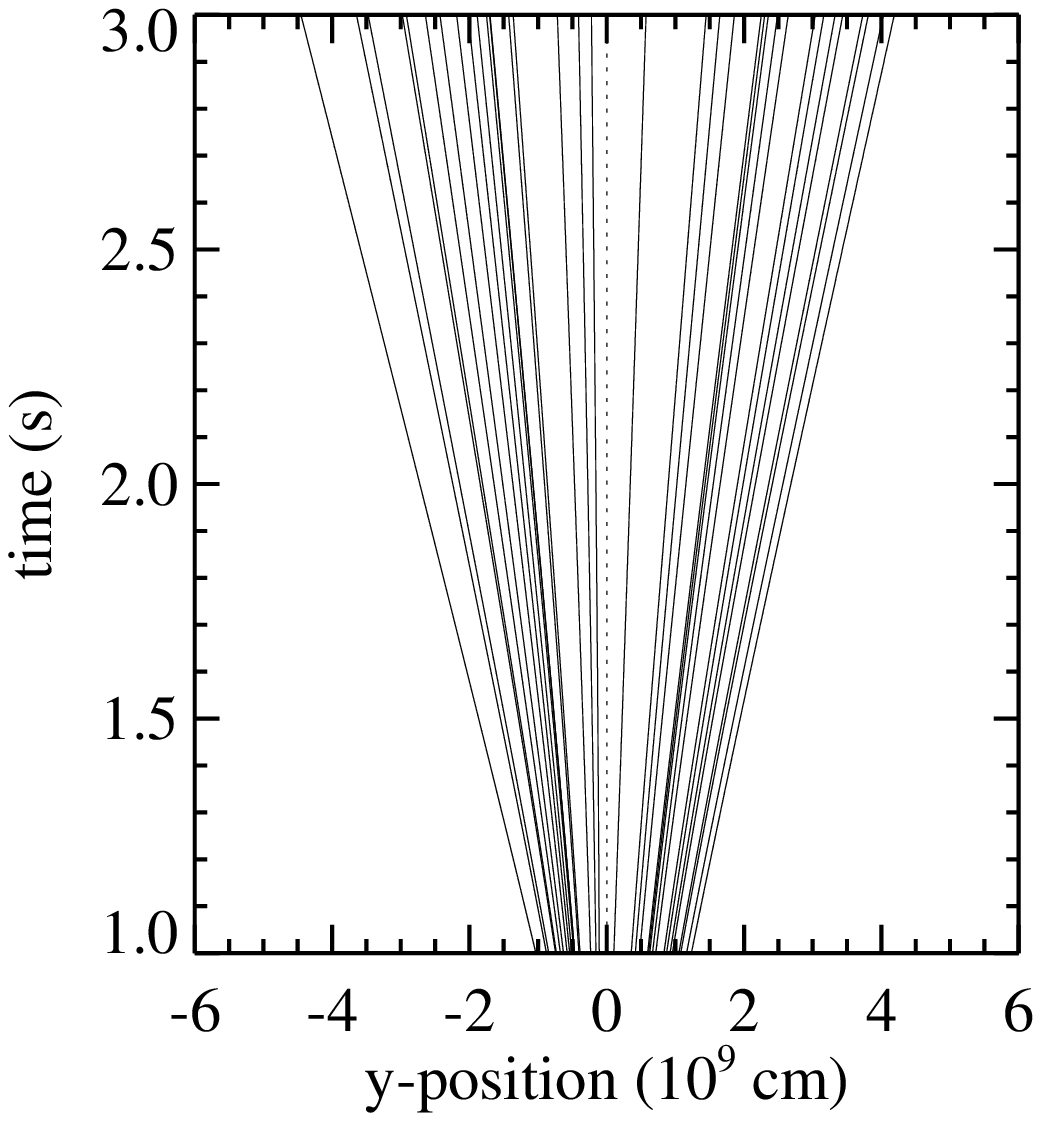}
  \caption{(left) On-axis Lagrangian tracer particle positions are
    shown as a function of time in this space-time diagram.  The thick
    dashed line shows the theoretical position of a constant speed
    detonation originating from the location $y=0.2\ee{9}\nsp\cm$,
    coincident with the detonation initiated in the simulation.  A
    theoretical detonation speed taken to be $v_{\rm det} =
    1.1\ee{9}\nsp\cmpersec$ for the detonation as it moves inward
    toward the high density core and a speed that is 10\% lower as it
    moves outward into the low density surface material provides a
    very good match to the simulation data.  (right) At times greater
    than $1\second$ the Lagrangian tracer particles exhibit homologous
    expansion.}
  \label{f:SpaceTimeTrajs}
\end{figure*}

A revealing format for presenting the dynamics of the detonation and
the subsequent expansion is the space-time diagram. In
Fig.~\ref{f:SpaceTimeTrajs} we present the space-time trajectories for
all of the tracer particles that were initialized near the symmetry
axis of the white dwarf. The left panel shows the time period over
which the detonation traverses the stellar core, while in the right
panel we show the later time evolution that ends in a radially
expanding, ballistic trajectory for each of the particles.  The bold
dashed line in the left panel shows the path taken by a theoretical
detonation having a constant speed, which matches the kinks in the
particle trajectories very well.

\begin{figure}[htbp]
  \centering{\includegraphics[width=3.4in]{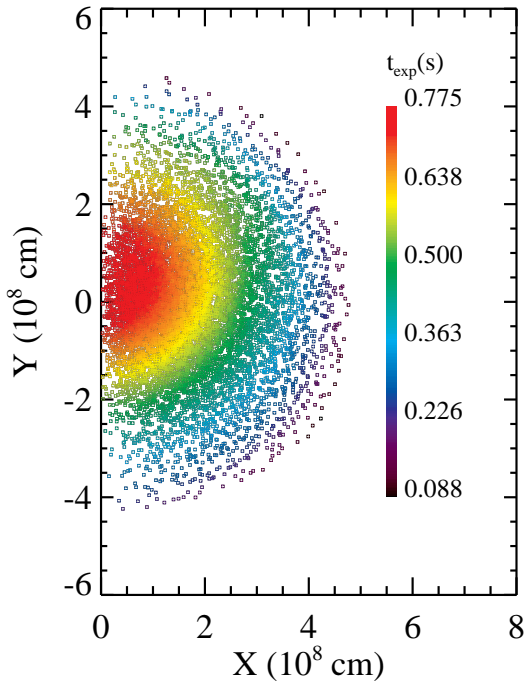}}
  \caption[Expansion time scale as a function of initial
  position]{Initial spatial position of the Lagrangian tracer particles. Color
    represents each particle's expansion time scale. The center of the
    star is at X=0 Y=0 The detonation was initiated at X=0
    Y=2. Particles on the side of the star where the detonation starts
    have higher expansion time scales than the particles on the
    opposite side of the star. The expansion time scale is calculated from the temperature profile of each tracer.}
  \label{f:SpaceDist}
\end{figure}

The following features in this figure merit further discussion. (1)
The trajectories are slowly converging prior to detonation. This is
the signature of the stellar core undergoing mild contraction as a
result of having been expanded by a radial pulsation mode prior to
detonation. (2) The detonation accelerates material in the direction
it is propagating.  This is the primary source of the asymmetry
imprinted on the remnant at late times.  A large number of Lagrangian tracer
particles in the detonated hemisphere are first accelerated towards
the stellar center by the detonation before they are turned around by
pressure forces and accelerated to their final, outwardly directed
expansion velocities. The exact number of tracer particles accelerated towards
the stellar center by the detonation is dependent on distance from the center of the star to the point where the detonation was initiated, $a$. 
A given tracer particle with a central angle $\theta$ and distance from the center of the star $r$ will be accelerated towards the center of the star if
$sin(\theta) > r/a$.  
On the other hand, tracers in the opposite
hemisphere are accelerated by the detonation in the same direction as
their final expansion velocity, reaching their final velocity on a
shorter timescale. (3) The material in the detonated hemisphere is
accelerated to lower velocities overall compared to material in the
opposite hemisphere (see Fig.~\ref{f:SpaceTimeTrajs} (right)). This
mapping between initial position (and therefore initial density) and
resultant expansion velocity explains the density profile: the
material lines in the more rapidly expanding hemisphere are stretched
out over a larger region of space, and hence to a lower relative
density, than the more slowly expanding regions. (4) A natural
consequence of the explosion dynamics is an asymmetry in the expansion
timescale $t_{exp}$, defined as the time required for the detonated
material to drop from its post-detonation temperature maximum
$T_{max}$ to $e^{-1}T_{max}$, resulting from the differential rate at
which material cools (nearly adiabatically) due to the post-detonation
expansion. This follows directly from point (2) above. The expansion
timescale asymmetry is shown in Fig.~\ref{f:SpaceDist} where we have
plotted the tracer particles at their initial position in the stellar
core, color coded by their post-detonation expansion timescale.  It is
obvious from this figure that the material in the detonated hemisphere
$(y>0)$ has overall a larger expansion timescale than in the opposite
hemisphere for a given initial upstream density. As will be discussed
in \S\S\ref{s:NucDepend} and \ref{s:Phenom} this expansion timescale
distribution imparts an asymmetry in the resultant nucleosynthetic
yield.

\subsection{Nucleosynthesis Dependence}\label{s:NucDepend}

We find that nuclear burning in \SNeIa\ progresses in three distinct
stages \citep{Khokhlov1983Deflagration-Fr,khokhlov1991}. The first
stage is carbon burning.  During carbon burning \carbon+\carbon\ is
the primary reaction taking place. We find that carbon burning never
reaches an equilibrium state in a small region of the star where the
final carbon mass fraction is above $10^{-4}$. The carbon burning
reactions are sensitive to temperature with
\begin{equation}\label{eq:dYdt}
  \frac{dY(\carbon)}{dt} \propto f(T_9) T_9^{-2/3} e^{-84.165 T_9^{-1/3}}.
\end{equation} 
Here $dY(\carbon)/dt$ is change in \carbon\ abundance over change in
time, $T_{9}$ is temperature in units $10^{9}\nsp\K$, and $f(T_9)$ is
a function defining the temperature effect on the branching ratio
between the $ \carbon ( \carbon , \alpha ) \neon$ and $ \carbon (
\carbon , p ) \sodium[23]$ reaction \citep{caughlan88:_therm}.  Any
change in the thermal profile will result in a different abundance
pattern for material that does not proceed to the next phase of
burning.

The next stage is oxygen burning. Here \oxygen\ and the products of
carbon burning proceed to silicon group elements, like S, Ar, and Si.
As in carbon burning, oxygen burning never reaches an equilibrium state
and therefore also shows a dependence on the thermal history. We find
$X(\oxygen) > 10^{-5}$ if the next burning stage did not start. Very little
mass of the star ($<0.04\Msun$) is in a region that does not complete either carbon
or oxygen burning, so most of the star proceeds to the next burning
stage.

At higher temperatures and densities, silicon burning is the dominant
form of nucleosynthesis.  In silicon burning groups of nuclides enter
into equilibrium; a state known as quasi-statistical equilibrium
\citep[QSE; e.g.,][]{woosley1973}. There are two ways in which
changing the expansion time scale can affect the abundances in
QSE. First, while equilibrium holds within a group of nuclides, it
does not hold between groups.  Second, within a group of nuclides in
equilibrium reactions freeze out at different temperatures, resulting
in an abundance pattern that depends on the expansion time scale. This
will be further explored below. 

\begin{figure}[htbp]
  \centering{\includegraphics[width=3.4in]{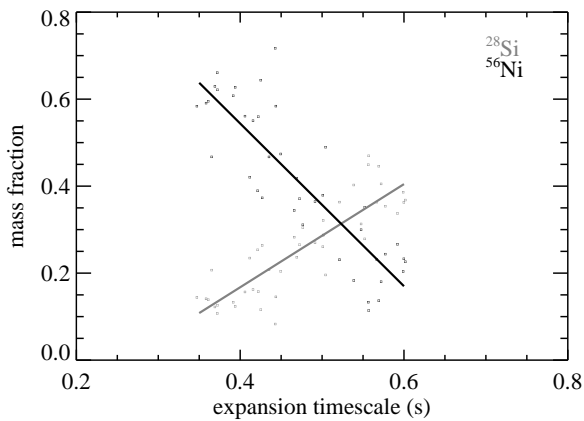}}
  \caption[Abundance as a function of expansion time scale]{Mass
    fraction of \silicon (red) and \nickel (blue) as a function of
    expansion time scale. This plot was made from tracer particles
    that had a maximum temperature between $4.99\ee{9}\nsp\K$ and
    $5.01\ee{9}\nsp\K$ and a density of approximately
    $1.5\ee{7}\nsp\grampercc$.The lines shown are least squares fits
    to the data.}
  \label{f:Yvstexp}
\end{figure}

\begin{figure*}[htbp]
  \includegraphics[width= 3.5in]{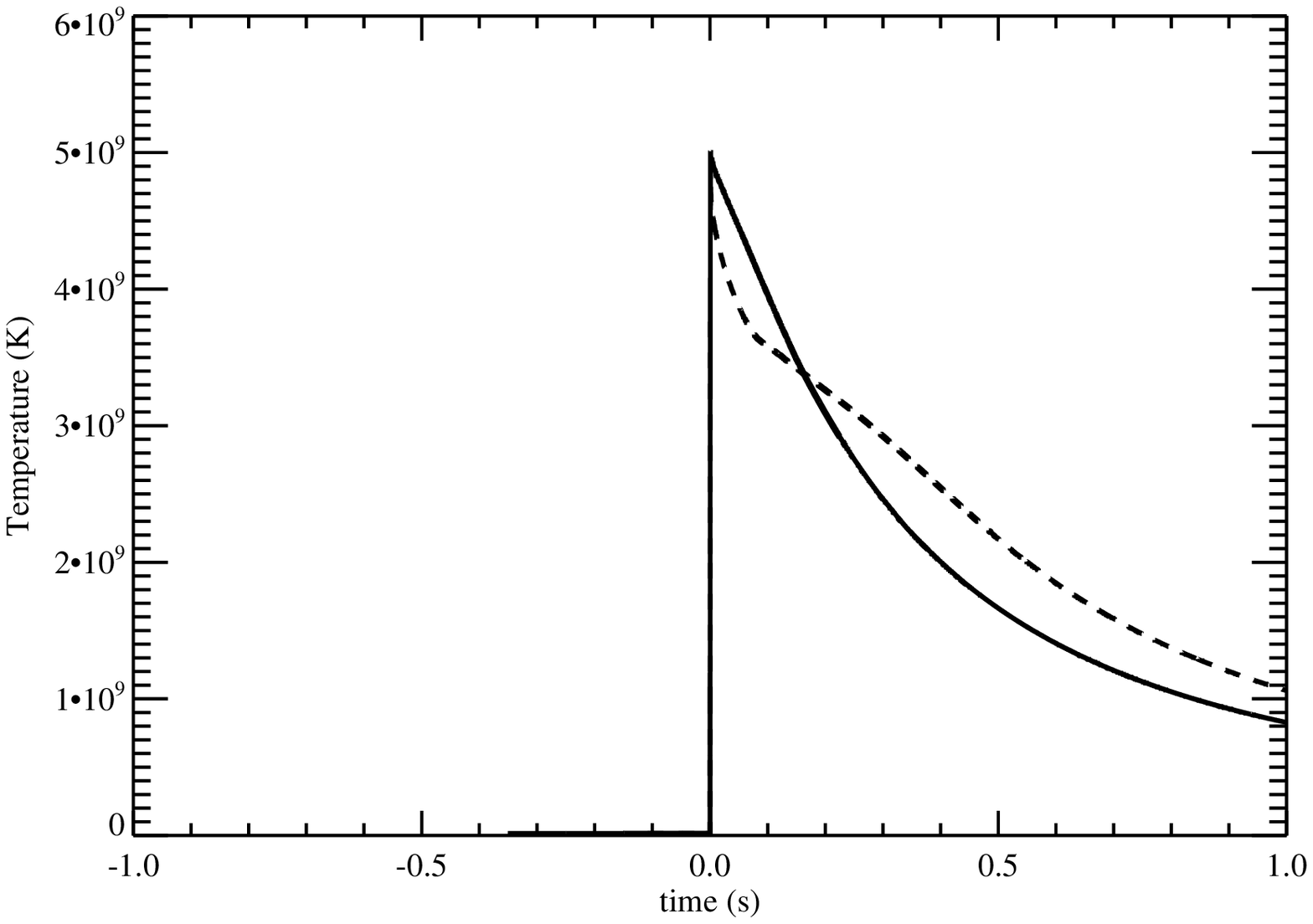}
  \includegraphics[width= 3.5in]{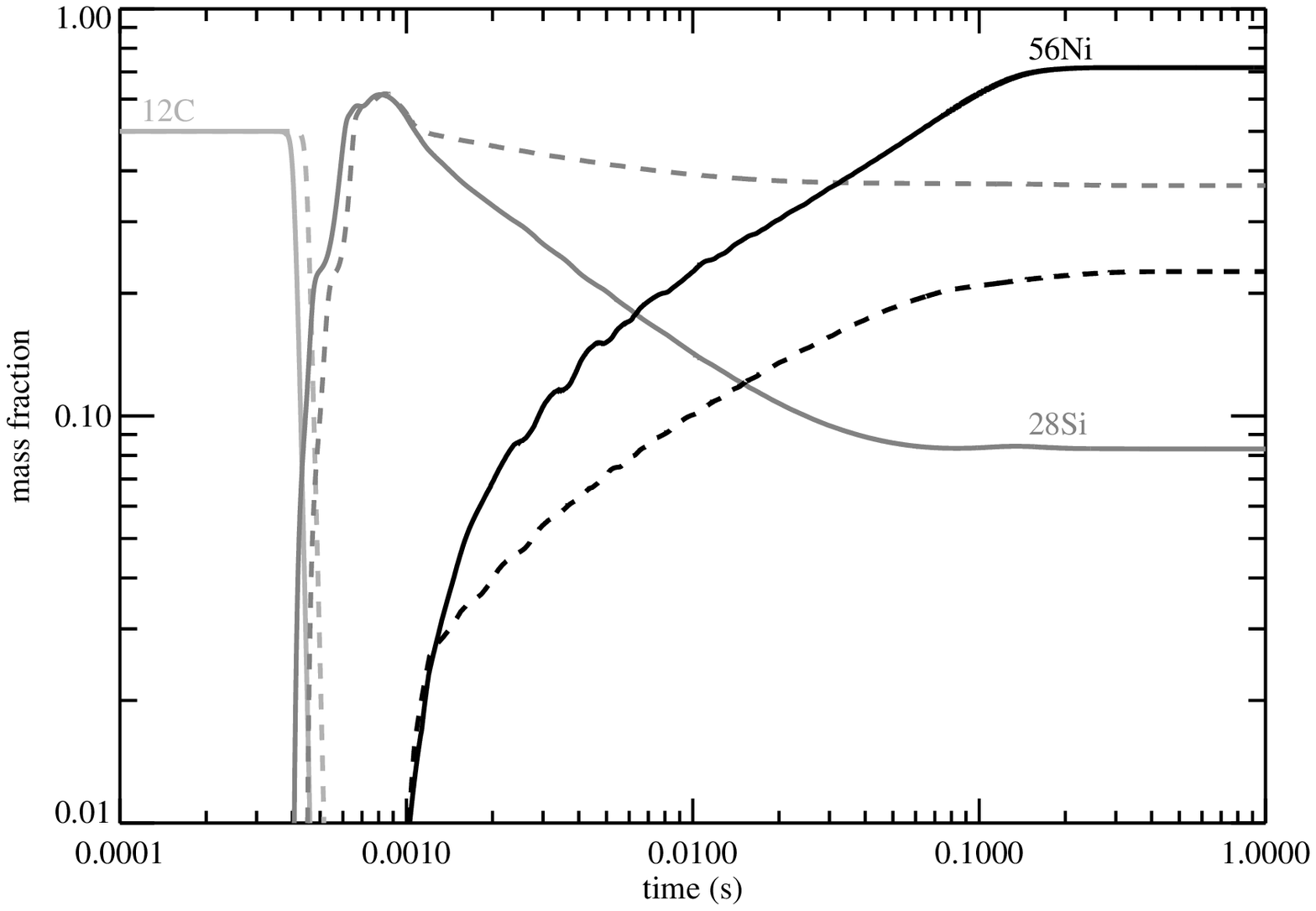}
  \caption{(left) Temperature profile for two particles with similar
    peak temperatures but different expansion time scales. The
    expansion time scale for the solid line is $0.4428\nsp\second$ and
    the expansion time scale for the dashed line is
    $0.6016\nsp\second$. 
    The times for both particles have been offset
    such that the peak temperature is reached at
    $1\ee{-3}\nsp\second$. (right) The mass fraction of \carbon,
    \silicon, and \nickel\ as a function of time for the same
    time-adjusted particles.}
  \label{f:TComp}
\end{figure*}

Fig.~\ref{f:Yvstexp} shows how the abundances of silicon and nickel
vary as functions of expansion time scale over a small range of
temperatures for tracer particles that never finish silicon
burning. Even with the scatter from plotting particles with different
peak temperatures, there is a clear dependence of the abundances on
expansion timescale. The directions of these trends are
counterintuitive, but Fig.~\ref{f:TComp} shows their origin. Both
tracer particles shown reach a peak temperature of $\sim
5\ee{9}\nsp\K$, and their nucleosynthesis is nearly identical up to
that point. The particle with the longer $(0.6016\nsp\second)$ thermal
expansion time scale, however, has a temperature that falls faster
over the first $0.1\nsp\second$ than the particle with the shorter
$(0.4428\nsp\second)$ thermal expansion time scale. This rapid decrease
in temperature results in less \silicon\ burned to \nickel[56] even
though the thermal expansion time scale is longer.

Material exposed to high enough density and temperature conditions for
long enough will arrive in a state of nuclear statistical equilibrium
(NSE). In this state all nuclear reactions enter equilibrium and lose
all history of the thermal evolution up to that point.  For material
that has reached this state, the only dependence that the final yield
has on expansion timescale occurs during the process of freeze
out. Freeze out occurs for a nuclide when the temperature drops low
enough that all strong reactions become too slow to change the
nuclides abundance again. Because this condition occurs at different
temperatures for different nuclides, the final yield depends on the
rate at which material cools. Like in QSE, different reactions freeze out
at different temperatures, leading to yields that depend on the
expansion timescale. Therefore, all three stages of burning and NSE in
\SNeIa\ are affected by different expansion time scales,  resulting
in a clear compositional asymmetry that will be discussed in the
following section.

\subsection{Phenomenology}\label{s:Phenom}
We now present the results of our reaction network calculations for a
near edge-lit detonation in a \SNIa\ model. We find that a number of
nuclides exhibit pronounced asymmetries across the stellar remnant. We
quantify this effect by calculating the center of mass for a given
element.  Suppose a tracer particle $i$ has a mass fraction $X_i(Z)$
of element $Z$, and its position is given by vector
$\mathbf{r}_i$. Then the center of mass for a given element,
$\mathbf{r}_{cm}(Z)$, is given by the equation
\begin{equation}\label{eq:cm}
  \mathbf{r}_{cm}(Z) = \left(\sum_{i} X_i(Z) \mathbf{r}_i\right) / \left(\sum_{i} X_i (Z)\right).
\end{equation} 
Due to the cylindrical symmetry of our explosion model the
displacement of the center of mass for any element lies along the y
axis. A velocity for the center of mass $\mathbf{v}_{cm}$ can be
calculated by replacing $\mathbf{r}_i$ by the velocity $\mathbf{v}_i$
in equation \ref{eq:cm}.  Table \ref{t:gradtab} shows the total mass,
displacement of the center of mass, and velocity of the center of mass for elements between carbon and
germanium. For reference, the detonation was initiated at $\sim
2\ee{8}\nsp\cm$. These numbers are correct for the end of our
simulation $(t \sim 3\nsp\second)$, where strong reactions have frozen out and homologous
expansion has been reached. Some of the isotopes making these
elements, \nickel[56] for example, decay so Table \ref{t:gradtab}
evolves with time. The last three columns show the total mass, displacement of the center of mass, and velocity of the center of mass assuming all radioactive elements 
instantly decayed to their stable isotopes.
Note that nickel shows no change in center of mass or velocity since most of the mass of the star ends up as radioactive nickel in our model.
If the complete star was burned to nickel then by definition the change in center of mass would be zero since our model conserves mass and momentum. 
Elements lighter than silicon have their masses
distributed more in the direction where the detonation was
initiated. Elements heavier than silicon are, for the most part,
distributed away from where the detonation was initiated. These
elements also display an odd-even pattern where odd Z nuclei, like cobalt and copper, are
predominantly distributed farther away from the start of the
detonation than their even Z counterparts, like iron and zinc. This asymmetry is due to the different thermal histories of the two sides of the star affecting the nucleosynthesis as outlined in \S\S\ref{s:NucDepend}.

\begin{deluxetable*}{ccccccc}
  \tablecaption{Centers of mass and velocities for various elements}
  \tablehead{ & \colhead{$\Delta_{cm}(\sim3\nsp\second)$} & \colhead{$V_{cm}(\sim3\nsp\second)$}  & \colhead{mass$(\sim3\nsp\second)$} & \colhead{$\Delta_{cm}(decayed)$} & \colhead{$V_{cm}(decayed)$}  & \colhead{mass$(decayed)$}\\
    & $(10^8\nsp\cm)$ & $(10^8\cmpersec)$ & $(\gram)$ & $(10^8\nsp\cm)$ & $(10^8\cmpersec)$ & $(\gram)$}
  \startdata
  C    &    29.1  &  9.9  & 9.00\ee{29} &  29.1 &  9.9 & 9.00\ee{29} \\
  N    &    18.2  &  6.2  & 2.91\ee{25} &  11.6 &  4.1 & 8.41\ee{26} \\
  O    &    7.30  &  2.4  & 7.28\ee{31} &  7.30 &  2.4 & 7.28\ee{31} \\
  F    &    27.9  &  9.4  & 2.19\ee{22} &  35.2 &  12. & 3.34\ee{23} \\
  Ne   &    28.7  &  9.9  & 1.03\ee{30} &  28.7 &  9.9 & 1.03\ee{30} \\
  Na   &    25.6  &  8.8  & 1.12\ee{27} &  22.3 &  7.6 & 5.69\ee{27} \\
  Mg   &    13.8  &  4.9  & 1.37\ee{31} &  13.8 &  4.9 & 1.37\ee{31} \\
  Al   &    21.6  &  7.5  & 1.74\ee{28} &  18.5 &  6.4 & 3.13\ee{28} \\
  Si   &   -1.09  & -0.6  & 2.18\ee{32} & -1.09 & -0.6 & 2.18\ee{32} \\
  P    &   -4.44  & -1.9  & 8.34\ee{28} &  3.77 &  1.2 & 4.87\ee{28} \\
  S    &   -2.68  & -1.2  & 1.18\ee{32} & -2.68 & -1.2 & 1.18\ee{32} \\
  Cl   &   -9.31  & -3.7  & 1.82\ee{28} & -4.28 & -1.8 & 1.21\ee{28} \\
  Ar   &   -2.95  & -1.3  & 2.60\ee{31} & -2.94 & -1.3 & 2.60\ee{31} \\
  K    &   -4.53  & -1.7  & 5.12\ee{27} & -4.20 & -1.7 & 3.66\ee{27} \\
  Ca   &   -2.55  & -1.1  & 2.80\ee{31} & -2.56 & -1.1 & 2.80\ee{31} \\
  Sc   &   -3.36  & -1.3  & 2.02\ee{26} & -9.21 & -3.7 & 1.57\ee{25} \\
  Ti   &   -2.78  & -1.1  & 3.93\ee{28} & -2.35 & -1.0 & 7.54\ee{29} \\
  V    &   -2.04  & -0.7  & 3.68\ee{27} & -3.91 & -1.5 & 2.45\ee{28} \\
  Cr   &   -2.37  & -1.0  & 7.60\ee{29} & -1.81 & -0.8 & 1.71\ee{31} \\
  Mn   &   -3.68  & -1.4  & 2.58\ee{28} & -6.01 & -2.4 & 2.91\ee{29} \\
  Fe   &   -1.80  & -0.8  & 1.71\ee{31} &  0.00 &  0.0 & 2.16\ee{33} \\
  Co   &   -5.59  & -2.2  & 3.06\ee{29} & -3.49 & -1.3 & 2.57\ee{30} \\
  Ni   &    0.00  &  0.0  & 2.16\ee{33} &  0.13 &  0.1 & 3.52\ee{31} \\
  Cu   &   -3.06  & -1.1  & 4.68\ee{30} & -2.08 & -0.7 & 2.46\ee{28} \\
  Zn   &    0.17  &  0.1  & 2.93\ee{31} & -1.49 & -0.5 & 6.46\ee{29} \\
  Ga   &   -2.09  & -0.7  & 3.09\ee{28} & -0.83 & -0.3 & 3.54\ee{25} \\
  Ge   &   -1.48  & -0.5  & 6.31\ee{29} & -0.47 & -0.1 & 3.11\ee{26} \\
  \enddata
  \label{t:gradtab}
\end{deluxetable*}

\begin{figure}[htbp]
  \centering{\includegraphics[width= 3.0in]{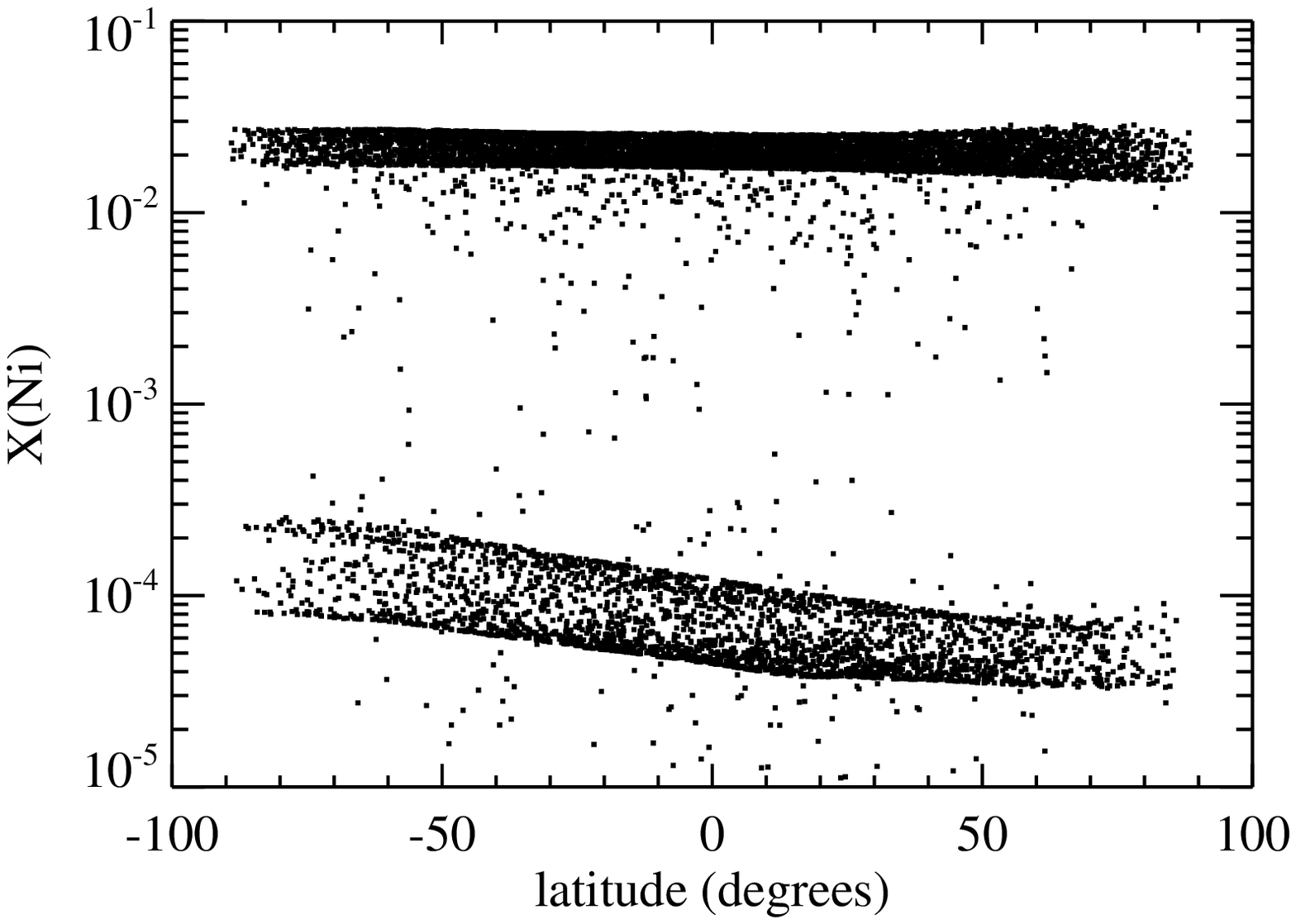}}
  \caption[Final mass fraction of elemental Ni as a function of central to 
  angle]{Final mass fraction of elemental(nonradioactive) Ni as a function of the ejection
    angle relative to the center of the star. The detonation started
    in the surface layer of the star in the theta = $90^{\circ}$ direction. The
    particles with a Ni mass fraction above $10^{-3}$ are particles that
    have burned to NSE.}
  \label{f:sample-abun}
\end{figure}

Our simulation was not run sufficiently past freeze out to allow all
beta unstable nuclei to decay. However, if we decay the unstable
nuclei and group them in elemental abundances, we find elemental nickel to
have a clear gradient over the star. In material not burned to
NSE we find the mass fraction of elemental nickel between $+90^\circ$
and $-90^\circ$ central angle to each tracer particle to increase by a factor of 2--3. Figure
\ref{f:sample-abun} shows the dependence of elemental nickel on central angle to each tracer particle.

\begin{figure}[htbp]
  \centering{\includegraphics[width=3.4in]{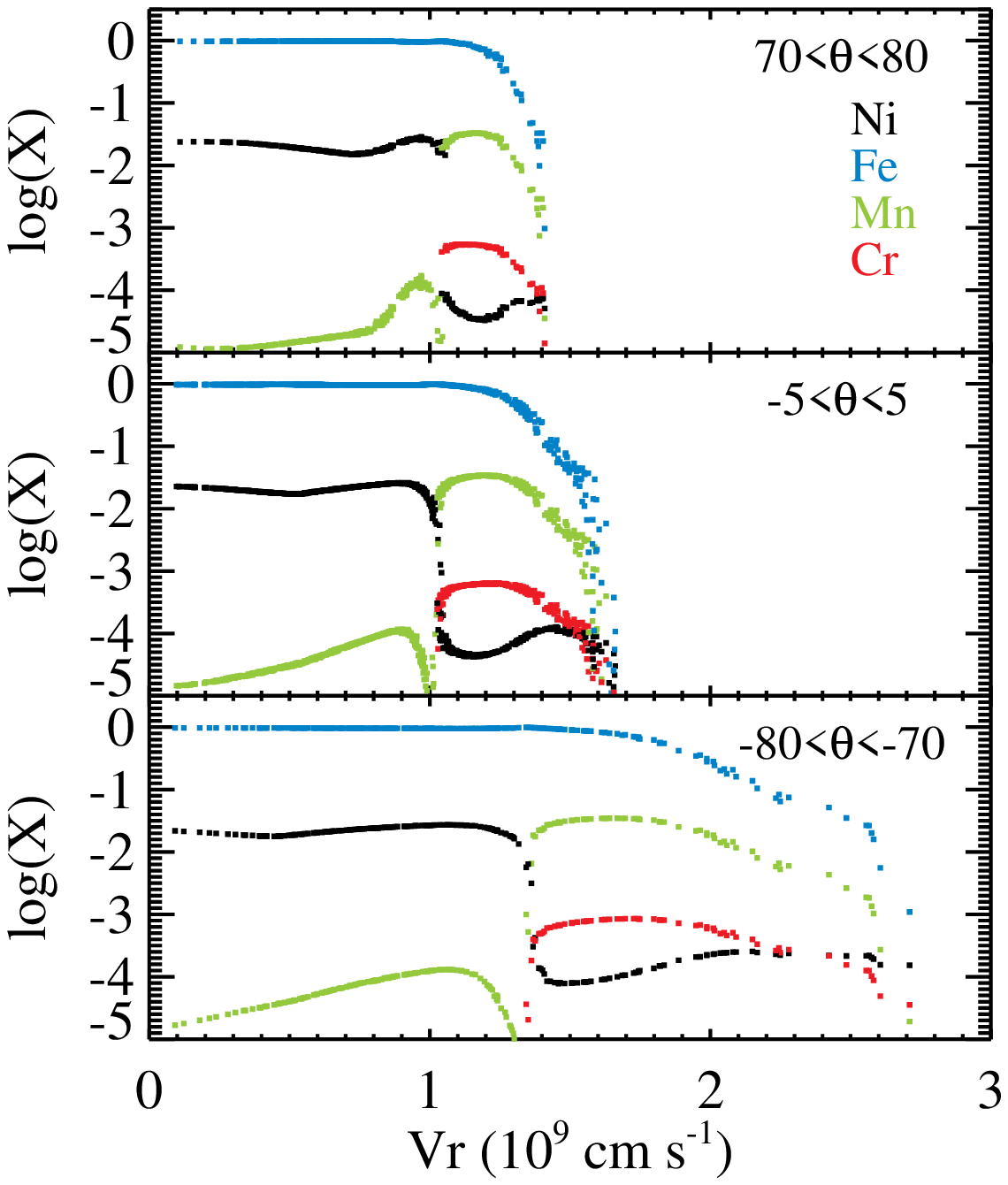}}
  \caption[Final mass fraction of elemental iron group elements as a function of radial
  velocity]{Final mass fraction of elemental iron group elements as a function of the final
    radial velocity. The particles with a Ni mass fraction above $10^{-3}$
    are particles that have burned to NSE. Particles ejected between $-80^{\circ}<\theta<-70^{\circ}$ reach the highest radial velocities.}
  \label{f:sample-abun2}
\end{figure}

An interesting side effect of the different expansion times is that
material on opposite sides of the star expands at different
velocities. This leads to a gradient in velocity space. Figure
\ref{f:sample-abun2} shows how elemental nickel, iron, manganese, and chromium vary with radial
velocity for different central angles. In material not burned to NSE, the part of the
remnant with the most nickel is also the part with the highest radial
velocity. This is self-consistent since the nickel mass fraction is greater on
the side of the remnant with the shortest expansion time. Therefore, it
follows it should have the highest radial velocity.

\section{Discussion}
\label{s:discussion}
We have computed the abundances and spatial distributions of nuclides
in an explosion of an expanded near Chandrasekhar-mass white dwarf resulting
from an off-center initiated detonation, a toy model that captures the
thermodynamic profile of some \SNIa\ explosion models.  We find a compositional
asymmetry in the ejecta produced by the detonation. This compositional
asymmetry is connected with the thermal expansion time scale. The
different expansion timescales also result in a compositional
asymmetry in velocity space.

It is difficult to establish the observable features of our model since we have not conducted radiative transfer calculations to generate light curves and spectra. It is currently unclear how much of an observational effect this asymmetry will have. A series of synthetic spectra generated over a range of time allows for direct comparison with observed supernovae. We conjecture, that even if compositional effects are obscured the spectra will show some dependence on observing angle. This is because the side of the remnant that expands at higher velocities will also be at a lower density, making it more transparent at earlier times.

Even though it is difficult to determine observational features of our model, it is instructive to compare and contrast our model with other recent results. The `toy model' in \citet{Hillebrandt2007} and
\citet{Sim2007} is constructed similarly to our calculated yield distribution. They found that off-center distributions of burned material are likely to leave detectable imprints on observed light curves. An angular dependence of the light-curve peak brightness is introduced that might explain some over luminous \SNeIa . 

Our model has no deflagration ash so comparisons with deflagration to detonation models, like \citet{Livne1999}, is problematic. 
We can compare to \citet{maeda2010}, who showed that expansion velocity gradients as
inferred from the Si II $\lambda6355$ absorption feature could be
explained by a velocity shift of $3500\nsp\km\usp\second^{-1}$ in Si
happening in an `opening angle' $105-110^\circ$ away from the ignition
points caused by deflagration ash. Even though our model has no deflagration ash 
a similar effect occurs with \silicon\ having ejection
velocities $\sim10000\nsp\km\usp\second^{-1}$ faster on the side of
the star opposite of the detonation. The ejecta in our model, however,
do not have a sharp transition in velocity but a gradual increase in
velocity starting at a point $\sim90^\circ$ away from the ignition
point. It is difficult to determine if our model can reproduce the
observed velocity gradients. It is worth noting that the deflagration
to detonation transition model used in \citep{maeda2010} did not
produce the observed velocity shift or `opening angle' either.

\citet{kasen2005} attempted to calculate the spectral signatures of GCD by considering ejecta interacting with an extended atmosphere. The ejecta in their model were calculated from a 1-D simulation, and therefore lack the asymmetry in the nucleosynthesis that the reaction  network calculations of our 2-D simulation find. In \citep{kasen2007} nucleosynthesis was done approximately with a 13 element  reaction network. The surface flow, which consists partly of deflagration ash, which was excluded from our present model, needs to be considered. The surface flow might also have a spectral signature itself such as the presence of an high-velocity calcium absorber \citep{kasen2005} and should be compared with the underlying compositional asymmetry. In the case of
sub-Chandrasekhar models, detonation of a pure helium shell leads to a layer containing iron-group elements such as titanium and chromium around the core ejecta \citep{fink2010,sim2010,kromer2010}.

Another item to consider is that we do not include the effects of metallicity on the
nucleosynthesis. Our model is initially composed of \carbon\ and
\oxygen . However, it has been shown that prior to the explosion of a
carbon-oxygen white dwarf in a \SNIa\ there is a long period during
which some \carbon\ is converted into \carbon[13] as well as heavier
elements \citep{Chamulak2007The-Reduction-o}. This process makes even
the most metal poor \SNeIa\ have a composition of more diverse than pure \carbon\
and \oxygen. It is worth mentioning that for deflagration--detonation
transition (DDT) models where the detonation density was allowed to
vary in relation to the flame speed as a function of metallicity
\citep{Chamulak2007The-Laminar-Fla} the yield of \nickel[56] produced
also varied with metallicity \citep{bravo2010,jackson2010}. DDT
models with varying metallicity and fixed detonation density, however,
showed little variation in outcome \citep{Townsley:2009jl}.

\acknowledgements

The authors wish to acknowledge that this work was supported by the US
Department of Energy, Office of Nuclear Physics, under contract
DE-AC02-06CH11357. This work was also supported by the NSF, grant
AST-0507456, by the \textbf{J}oint \textbf{I}nstitute for
\textbf{N}uclear \textbf{A}strophysics at MSU under NSF-PFC grant
PHY~0822648, by the DOE through grant 08ER41570, and by the Deutsche
Forschungsgemeinschaft via the Emmy Noether Program (RO 3676/1-1). The
software used in this work was in part developed by the DOE-supported
ASC / Alliance Center for Astrophysical Thermonuclear Flashes at the
University of Chicago. Simulations presented in this work were run at
the High Performance Computing Center at Michigan State University.

\bibliographystyle{apj} \bibliography{master}

\end{document}